\documentclass[12pt,showpacs,preprintnumbers,pra,aps,amssymb,amsfonts]{revtex4}

\usepackage{times,graphicx}

\newcommand{\trace}{\mathop{\rm Tr}\nolimits}

\newcommand{\diag}{\mathop{\rm Diag}\nolimits}

\newcommand{\qed}{\hfill$\square$\par\vskip24pt}
\newcommand{\twomat}[4]{\left(\begin{array}{cc}#1&#2\\#3&#4\end{array}\right)}

\newcommand{\cT}{{\cal T}}

\newcommand{\R}{{\mathbf{R}}}

\newcommand{\id}{\mathbf{I}}

\newcommand{\be}{\begin{equation}}
\newcommand{\ee}{\end{equation}}
\newcommand{\bea}{\begin{eqnarray}}
\newcommand{\eea}{\end{eqnarray}}
\newcommand{\beas}{\begin{eqnarray*}}
\newcommand{\eeas}{\end{eqnarray*}}

\newtheorem{theorem}{Theorem}
\newtheorem{lemma}{Lemma}
\newtheorem{corollary}{Corollary}

\newtheorem{proposition}{Proposition}

\begin{document}
\title{Continuity bounds on the quantum relative entropy --- II}
\author{Koenraad M.R. Audenaert}
\email{koenraad.audenaert@rhul.ac.uk}
\affiliation{Department of Mathematics, Royal Holloway, University of London, Egham TW20 0EX, UK}
\author{Jens Eisert}
\affiliation{Dahlem Center for Complex Quantum Systems, Freie Universit\"at Berlin,
D-14195 Berlin, Germany}

\date{\today}

\begin{abstract}
The quantum relative entropy
is frequently used as a distance measure
between two quantum states, and inequalities relating it to other distance measures
are important mathematical tools in many areas of quantum information theory.
We have derived many such inequalities in previous work.
The present paper is a follow-up on this, and provides a sharp upper bound on the relative entropy
in terms of the trace norm distance and of the smallest eigenvalues of both states
concerned. The result obtained here is more general
than the corresponding one from our previous work.
As a corollary, we obtain a sharp upper bound on
the regularised relative entropy introduced
by Lendi, Farhadmotamed and van~Wonderen.
\end{abstract}
\maketitle

\section{Introduction}
The quantum relative entropy of states of quantum systems
is a measure of how well one quantum state can be operationally
distinguished from another and quantifies the extent to which one
hypothesis $\rho$ differs from an alternative hypothesis $\sigma$
in the sense of quantum hypothesis testing
\cite{HP,ohya,ON,Wehrl}.
It is defined as
\begin{eqnarray}\nonumber
    S(\rho||\sigma)=
    \trace\rho(\log \rho  - \log \sigma)
\end{eqnarray}
for states $\rho$ and $\sigma$ \cite{umegaki} 
whenever the support of $\rho$ is contained in the support of $\sigma$,
and is defined to be $+\infty$ otherwise.

In \cite{BAKA2} we presented a number of inequalities relating
the quantum relative entropy, used as a distance measure,
to the trace norm distance. The present paper is a follow-up on this work, and concerns a sharp upper bound
on the relative entropy $S(\rho||\sigma)$ in terms of the trace norm distance $||\rho-\sigma||_1/2$,
when the smallest eigenvalues of $\rho$ and $\sigma$ are given. The need for these smallest eigenvalues
stems from
the fact that the relative entropy can be infinite when the kernel of
$\sigma$ is not contained in the kernel of $\rho$.
Rastegin obtained similar inequalities for the relative $q$-entropy \cite{rastegin}.

As a special case of the main theorem proven here (Theorem \ref{th:main}),
we reobtain Theorem 6 of \cite{BAKA2}.
The proof given in \cite{BAKA2} was incorrect, and
the proof we give here serves as a correction and at the same time as a generalisation.

We also obtain an upper bound (Corollary \ref{cor:R}) on the so-called regularised relative entropy,
introduced by Lendi et.\ al.\ \cite{lendi}
as one possible means to circumvent the problem of infinities of the ordinary relative entropy.
The regularised relative entropy is defined as
$$
R(\rho||\sigma)=c_d\,\, S\left(\rho+\id || \sigma+\id\right),
$$
where $c_d$ is a certain normalisation constant depending on  $d$, the dimension of state space.
Note that $S\left(\rho+\id || \sigma+\id\right)\le \log 2$, with equality for orthogonal pure states, hence
one could also choose the normalisation constant to be $1/\log 2$.

In the following section, we introduce the notations and mathematical tools necessary for the proofs.
Then, in Section \ref{sec:rep}, we derive an integral representation for the relative entropy
between non-normalised states (i.e.\ positive definite matrices), which is also essential for the proofs.
An upper bound on the relative entropy for non-normalised states is derived in Section \ref{sec:main},
which is then used in Section \ref{sec:bounds} to obtain the promised
sharp upper bound on the relative entropy for normalised states.
\section{Notations and preliminaries}
In this paper we will work exclusively in finite dimensional Hilbert spaces, so that quantum states
can be represented by positive semidefinite matrices.
We denote the identity matrix by $\id$,
and scalar matrices $a\id$ simply by $a$ (for $a\in\R$) when no confusion can arise.
The matrix norms $||\cdot||_1$ and $||\cdot||_\infty$ are the trace norm and operator norm, respectively.

The von Neumann entropy can be defined for positive definite matrices as
\be
S(A) = -\trace A\log A,\label{eq:ent}
\ee
which coincides with the usual definition for density matrices.
Furthemore, we define $S(0)=0$.

Likewise, the quantum relative entropy can be defined for positive definite matrices $A$ and $B$ as
\be
S(A||B) = \trace A(\log A-\log B).\label{eq:relent}
\ee
This definition still holds for positive semidefinite $A$ and $B$ provided the support of $B$
is contained in the support of $B$; otherwise one defines $S(A||B)=+\infty$.
The quantum relative entropy satisfies the scaling property
\be
S(aA||aB)=aS(A||B), a>0.\label{scaling}
\ee

The logarithm appearing in (\ref{eq:ent}) and (\ref{eq:relent}) is the matrix logarithm.
For $x>0$, we have the following integral representation for the scalar logarithm:
\be
\log x = \int_0^\infty ds \left(\frac{1}{1+s}-\frac{1}{x+s}\right).
\ee
and for $A>0$ we define the matrix logarithm as
\be
\log A = \int_0^\infty ds \left(\frac{1}{1+s}-(A+s)^{-1}\right).\label{eq:intlog}
\ee

The methods we will use  require
the derivative of the matrix logarithm; see also \cite{KA1,KA2}.
From the integral representation of the logarithm we get, for $A>0$,
$$
\frac{d}{dt}\Bigg|_{t=0} \log(A+t\Delta)
=\int_0^\infty ds\,\,(A+s)^{-1} \Delta (A+s)^{-1}.
$$
As is customary, we define the following linear map for $A>0$:
\be
\cT_A(\Delta) = \int_0^\infty ds\,\,(A+s)^{-1} \Delta (A+s)^{-1}.\label{eq:tad}
\ee
Thus
\be
\frac{d}{dt}\Bigg|_{t=0} \log(A+t\Delta) = \cT_A(\Delta).\label{eq:dlogdt}
\ee
Again, (\ref{eq:tad}) and (\ref{eq:dlogdt}) are also valid for $A\ge0$ provided
$\ker A\subseteq\ker\Delta$.

From the integral representation of $\cT$ it follows that, for any $A>0$, $\cT_A$ preserves the
positive semidefinite order: if $X\le Y$, then $\cT_A(X)\le\cT_A(Y)$.

For $x>0$, the integral $\int_0^\infty ds \,\,\,x/(x+s)^2$ is equal to $1$.
Hence, for $A>0$,
\be
\cT_A(A) = \int_0^\infty ds\, (A+s)^{-1}\,A \, (A+s)^{-1} = \id.\label{eq:intproj}
\ee

An argument that we will use frequently is the special structure of the Jordan decomposition
of a traceless Hermitian matrix.
Let $\Delta$ be Hermitian, and $\trace\Delta=0$.
The Jordan decomposition of $\Delta$ is given by $\Delta=\Delta_+ - \Delta_-$, where $\Delta_+$ and $\Delta_-$
are positive semidefinite and mutually orthogonal, i.e.\ $\Delta_+\Delta_-=0$.
We have $\trace\Delta = \trace\Delta_+ - \trace\Delta_-$, hence $\trace\Delta_+ = \trace\Delta_-$.
Thus
\be
||\Delta||_1 =\trace\Delta_+ + \trace\Delta_- = 2\trace\Delta_+ .
\ee
It will also be clear that $||\Delta||_\infty$ is bounded above by $\trace\Delta_+ $, and thus
\be
||\Delta||_\infty \le ||\Delta||_1 /2, \label{traceless}
\ee
whenever $\Delta$ is traceless and Hermitian.
\section{An integral representation of the relative entropy\label{sec:rep}}
In this section we derive an integral representation of the quantum relative entropy for
non-normalised states,
$$
S(A||B)=\trace A(\log A-\log B),
$$
in terms of a differentiable path $s\mapsto C(s)$, where $C(0)=A$, $C(1)=B$ and $C(s)>0$.

\begin{lemma}
Let $A\ge0$ and $B>0$.
Let $0\le s\le1$ and let $s\mapsto C(s)$ be a continuous, differentiable path joining $A$ and $B$ (that is,
$C(0)=A$ and $C(1)=B$) and for all $s\in(0,1)$, $C(s)>0$.
Then the relative entropy $S(A||B)$ has the following integral representation:
\be
S(A||B) =\trace(A-B)+\int_0^1 ds\; \trace\frac{dC}{ds}(\log B - \log C(s)).\label{eq:RErep}
\ee
\label{lem:RErep}
\end{lemma}
In particular, if the path is linear, $C(s)=(1-s)A+sB$,
then (\ref{eq:RErep}) becomes
\be
S(A||B) = \trace(A-B)+\trace(B-A)\log B - \int_0^1 ds\; \trace(B-A)\log(A+s(B-A)).
\ee
\textit{Proof.}
First we rewrite $S(A||B)$ as
\beas
S(A||B) &=& \trace A(\log A-\log B) \\
&=&\trace(B-A)\log B + (S(B)-S(A)).
\eeas
Both terms can be written as integrals. For the first term we have
$$
\trace(B-A)\log B = \trace \int_0^1 ds\,\, \frac{dC}{ds}\log B.
$$
The second term can be written as:
$$
S(B)-S(A)
=-\trace (B-A)-\int_0^1 ds\,\,\trace \frac{dC}{ds}\log C(s).
$$
This can be shown as follows.
Let $f(x)=-x\log x$ be the function defining the von Neumann entropy $S(A)=\trace f(A)$. As $f'(x)=-1-\log x$,
we have, for $0<s<1$,
\beas
\frac{d}{ds} S(C(s))
&=& \frac{d}{ds} \trace f(C(s)) \\
&=& \trace f'(C(s))\frac{dC}{ds} \\
&=& -\trace \frac{dC}{ds} - \trace \frac{dC}{ds}\log C(s).
\eeas
Integrating over $s$ in the interval $[0,1]$ yields, indeed,
\beas
S(B)-S(A)
&=& S(C(1)) - S(C(0)) \\
&=& \int_0^1 ds\,\frac{d}{ds} S(C(s)) \\
&=& -\int_0^1 ds\,\trace \frac{dC}{ds} - \int_0^1 ds\,\trace\frac{dC}{ds}\log C(s) \\
&=& -\trace(B-A)- \int_0^1 ds\,\trace\frac{dC}{ds}\log C(s).
\eeas

Finally, adding the two integral representations yields (\ref{eq:RErep}).
\qed
\section{Upper bound on the relative entropy for non-normalised states\label{sec:main}}
In this section, we prove the main technical proposition (Proposition \ref{prop:AIBI})
from which the promised sharp bound will follow.
The proposition provides an upper bound on the relative entropy for
non-normalised states $A$ and $B$ with equal trace, in terms of the trace
norm distance $T$ between $A$ and $B$,
and of the minimal eigenvalues of $A$ and $B$.

We will denote the lowest eigenvalue of $A$ by $\alpha$, and the lowest eigenvalue of $B$ by $\beta$.
First we establish the allowed range of $T$ in terms of $\alpha$ and $\beta$.
It turns out that the trace norm distance between $A$ and $B$
cannot be smaller than $|\alpha-\beta|$:
\begin{lemma}\label{lem:constraint}
Let $A,B$ be positive semidefinite $n\times n$ matrices with $\trace A=\trace B$,
and $\lambda_{\min}(A)=\alpha$ and $\lambda_{\min}(B)=\beta$.
Then $T = ||A-B||_1 /2\ge|\alpha-\beta|$.
\end{lemma}
\textit{Proof.}
We assume first that $\alpha\ge\beta$.
Let $\Delta:=A-B$ have Jordan decomposition $\Delta=\Delta_+ - \Delta_-$.
Since $\trace A=\trace B$, we have $\trace\Delta=0$, hence $||A-B||_1 = 2\trace\Delta_+$.

Denoting the vector of eigenvalues sorted in non-increasing order by the symbol $\lambda^\downarrow$,
we then clearly have
$$
||A-B||_1 = 2\trace\Delta_+\ge 2\lambda^\downarrow_1(\Delta_+) = 2\lambda^\downarrow_1(A-B).
$$
Now, by Lidskii's Theorem (e.g.\ inequality (III.12) in \cite{Bhatia}), for all Hermitian $A$ and $B$,
the vector $\lambda^\downarrow(A)-\lambda^\downarrow(B)$ is majorised by the vector $\lambda^\downarrow(A-B)$.
In particular,
$$
\lambda^\downarrow_1(A-B)\ge \max_j \{\lambda^\downarrow_j(A)-\lambda^\downarrow_j(B)\}
\ge \lambda^\downarrow_n(A)-\lambda^\downarrow_n(B).
$$
By the hypothesis of the lemma, the last expression is equal to $\alpha-\beta$.

Hence we have shown that $||A-B||_1 \ge 2(\alpha-\beta)$ when $\alpha-\beta\ge0$.
When $\alpha-\beta\le0$  we can just swap the roles of $A$ and $B$ and obtain
$||A-B||_1 \ge 2(\beta-\alpha)$.
\qed

Because of the scaling property (\ref{scaling})
we can restrict ourselves to the case $\beta=1$.

\begin{proposition}\label{prop:AIBI}
Let $A,B$ be positive definite with $\trace A=\trace B$,
$\lambda_{\min}(A)=\alpha$, $\lambda_{\min}(B)=1$
and $T:=||A-B||_1/2$. Then $T\ge|\alpha-1|$, and
Then, for $\alpha>0$,
\be
S(A||B) \le (1+T)\log(1+T)-\alpha\log(1+T/\alpha),\label{eq:AIBI1}
\ee
where $\alpha\mapsto -\alpha\log(1+T/\alpha)$ is monotone decreasing, and
$-\alpha\log(1+T/\alpha)=:0$ for $\alpha=0$.

Moreover, quality can be achieved for any allowed values of $\alpha$ and $T$.
\end{proposition}

The proof relies on the following lemma:
\begin{lemma}\label{lem:AB}
Let $a$ and $b$ be two positive definite matrices with $\trace a=\trace b$, and let
$t=||b-a||_1/2$.
If $a\ge\gamma$, with $\gamma$ a non-negative scalar, then
\be
\cT_{b}(b-a) \le \frac{t}{\gamma+t}.
\ee
\end{lemma}
\textit{Proof.}
Let $\delta=b-a$,
which is Hermitian with trace equal to $0$ and trace norm equal to $2t$.
Thus, by (\ref{traceless}), $||\delta||_\infty \le t$, or $\delta \le t$.
We also have $t\gamma \le ta$. Combining the two inequalities yields
$\gamma\delta\le ta=t(b-\delta)$.
Hence,
$$
b-a = \delta \le \frac{t}{\gamma+t}b.
$$
Since the operator $Y\mapsto \cT_X(Y)$ is order-preserving for $X>0$,
applying this operator to both sides yields
\beas
\cT_b(b-a) &\le& \frac{t}{\gamma+t} \cT_{b}(b)
= \frac{t}{\gamma+t}.
\eeas
\qed

\textit{Proof of Proposition \ref{prop:AIBI}.}
We consider strictly positive $\alpha$ first.
Let us apply Lemma \ref{lem:AB} to the case $a=A$ and $b=A+x(B-A)$,
with $A$ and $B$ the matrices of the proposition
and $0\le x\le 1$. Let $\Delta=B-A$. Then $\delta=x\Delta$, $t=xT$ and $\gamma=\alpha$.
By the lemma, we then have (after dividing by $x$)
\be
\cT_{A+x\Delta}(\Delta) \le \frac{T}{\alpha+xT}.\label{eq:up}
\ee
Likewise, by setting $a=B$, $b=B+(1-x)(A-B)$ and $\gamma=1$, we get
$$
\cT_{B+(1-x)(-\Delta)}(-\Delta) \le \frac{T}{1+(1-x)T}.
$$
Noting that $B-(1-x)\Delta=xB+(1-x)A=A+x\Delta$, this yields the lower bound
\be
\cT_{A+x\Delta}(\Delta) \ge \frac{-T}{1+(1-x)T}.\label{eq:lo}
\ee
Again we exploit the Jordan decomposition of $\Delta$,
$\Delta=\Delta_+ - \Delta_-$ with $\Delta_+,\Delta_-\ge0$ and $\trace\Delta_+ = \trace\Delta_-=T$,
due to the facts that $\trace\Delta=0$ and $||\Delta||_1=2T$.
Combining this with (\ref{eq:up}) and (\ref{eq:lo}), and exploiting the fact that
for $X\ge0$, $Y\le y$ implies $\trace XY\le y\trace X$, we get
\beas
\trace\Delta\cT_{A+x\Delta}(\Delta)
&=& \trace\Delta_+ \cT_{A+x\Delta}(\Delta) - \trace\Delta_- \cT_{A+x\Delta}(\Delta)\\
&\le& T\left(\frac{T}{\alpha+xT}-\frac{-T}{1+(1-x)T}\right).
\eeas
Now let $s$ be a scalar, $0\le s\le 1$.
Integrating the previous inequality over $x$ from $s$ to $1$ yields
\beas
\trace\Delta(\log(A+\Delta)-\log(A+s\Delta))
&\le& T(\log(\alpha+T)-\log(\alpha+sT)+\log(1+(1-s)T)).
\eeas
Integrating a second time, now over $s$ from $0$ to $1$, yields:
\beas
\int_0^1 ds\,\,\,\trace\Delta(\log B-\log(A+s\Delta))
&\le& (1+T)\log(1+T)+\alpha(\log\alpha-\log(\alpha+T)).
\eeas
To finish the proof, we define the rectilinear path $C(s)=sB+(1-s)A$, for which $dC/ds=B-A=\Delta$,
and use Lemma \ref{lem:RErep}
to show that the left-hand side is just $S(A||B)$.

The strict positivity of $\alpha$ is required to satisfy the conditions of
Lemma \ref{lem:AB}.
However, by continuity of the relative entropy in its first argument, the bound
must be valid for $\alpha=0$ too. In the limit of $\alpha$ tending to $0$,
$\alpha(\log\alpha-\log(\alpha+T))$ goes to $0$.

Finally, we show that equality can be obtained for every allowed value of $T$ and $\alpha$.
Indeed, taking
$$
A=\twomat{1+T}{0}{0}{\alpha} \mbox{ and }
B=\twomat{1}{0}{0}{T+\alpha}
$$
satisfies all the requirements of the proposition and yields equality in (\ref{eq:AIBI1}).
\qed

\section{Sharp upper bounds on the relative entropy and regularised relative entropy\label{sec:bounds}}

Proposition \ref{prop:AIBI} allows us to derive an upper bound on
the ordinary relative entropy between density operators $\rho$ and $\sigma$
when the eigenvalues of $\rho$ and $\sigma$
are bounded below by the values $\alpha$ and $\beta$, respectively.
\begin{theorem}\label{th:main}
Consider density matrices $\rho$ and $\sigma$, with smallest
eigenvalues
$\lambda_{\min}(\rho)=\alpha$ and $\lambda_{\min}(\sigma)=\beta$.
Then $T:=||\rho-\sigma||_1/2\ge|\alpha-\beta|$
and, for $\alpha,\beta>0$,
\be
S(\rho||\sigma) \le (\beta+T)\log(1+T/\beta) - \alpha\log(1+T/\alpha),\label{thebound}
\ee
and, in the limit $\alpha\to0$,
\be
S(\rho||\sigma) \le (\beta+T)\log(1+T/\beta).
\ee
\end{theorem}
\textit{Proof.}
We use the scaling property and Proposition \ref{prop:AIBI},
with $A=\rho/\beta$ and $B=\sigma/\beta$. The formula of Proposition \ref{prop:AIBI}
can be taken over completely
by replacing $\alpha$ by $\alpha/\beta$, $T$ by $T/\beta$,
and multiplying the right-hand side of each bound by $\beta$.
\qed

Note that, because of the extra normalisation requirement $\trace\rho=\trace\sigma=1$,
equality can now only be achieved for states of dimension at least $3$.

If $\alpha$ is not specified, we must take the maximum of $(\beta+T)\log(1+T/\beta) - \alpha\log(1+T/\alpha)$
over all allowed values of $\alpha$, with $\beta$ and $T$ kept fixed.
In doing so we retrieve Theorem 6 of \cite{BAKA2}. The proof given here supplants the
incorrect proof in the published version of \cite{BAKA2}.
\begin{corollary}\label{cor:main}
Consider density matrices $\rho$ and $\sigma$, where
$\sigma$ has smallest eigenvalue $\lambda_{\min}(\sigma)=\beta$. Let $T:=||\rho-\sigma||_1/2$.
If $T\le\beta$
\be
S(\rho||\sigma) \le (\beta+T)\log(1+T/\beta) + (\beta-T)\log(1-T/\beta),
\ee
and if $T\ge\beta$,
\be
S(\rho||\sigma) \le (\beta+T)\log(1+T/\beta).
\ee
\end{corollary}
\textit{Proof.}
Let $\lambda_{\min}(\rho)=\alpha$. To find an upper bound on $S(A||B)$
in the case that $\alpha$ is not specified,
we maximise the bound (\ref{thebound}) over all allowed $\alpha$.
Because of Lemma \ref{lem:constraint}, $T\ge|\alpha-\beta|$. Hence, the range of $\alpha$
is $[\max(0,\beta-T) , \beta+T]$.
The quantity to be maximised is $- \alpha\log(1+T/\alpha)$, which is monotonously decreasing in $\alpha$.
Thus, its maximum is attained for the minimally allowed $\alpha$, being $\max(0,\beta-T)$.
The two cases of the corollary follow.
\qed

We immediately obtain an upper bound on the
regularised relative entropy in terms of the trace norm distance.
\begin{corollary}\label{cor:R}
For $d$-dimensional density matrices $\rho$ and $\sigma$, with smallest
eigenvalues
$\lambda_{\min}(\rho)=\alpha$ and $\lambda_{\min}(\sigma)=\beta$
and $T:=||\rho-\sigma||_1/2$,
\bea
R(\rho||\sigma)&:=&c_d\,\, S\left(\rho+\id||\sigma+\id\right) \\
&\le& c_d\;\left((\beta+1+T)\log(1+T/(\beta+1)) -(\alpha+1)\log(1+T/(\alpha+1))\right) \label{eq:Q} \\
&\le& c_d\; T\log(1+T).\label{eq:Q2}
\eea
From dimension 3 onwards, inequality (\ref{eq:Q2}) is sharp. 
Equality can be achieved for any allowed value of $T$, by the diagonal states
$\rho=\diag(1-t,t,0)$ and $\sigma=\diag(1-t,0,t)$, where $t$ can be any number between 0 and 1.
\end{corollary}
\textit{Proof.}
Upper bound (\ref{eq:Q}) is a straightforward application of Theorem \ref{th:main} (apart from a rescaling of
$\rho+\id$ and $\sigma+\id$, which has no effect on the bound itself).
When no information about $\alpha$ and $\beta$ is known one can use the bound (\ref{eq:Q2})
which follows by exploiting the fact that both $(\beta+1+T)\log(1+T/(\beta+1))$
and $-(\alpha+1)\log(1+T/(\alpha+1))$ are monotonically decreasing, hence expression (\ref{eq:Q}) is
maximal for $\alpha=\beta=0$.
\qed
\begin{acknowledgments}
This work was supported by
the European Commission (Qessence, Compas, Minos)
and the European Research Councils (EURYI).
We gratefully acknowledge the referee for many invaluable comments.
\end{acknowledgments}

\end{document}